\documentclass{article}
\usepackage{graphicx} 
\usepackage{amsmath}
\usepackage{amssymb}
\usepackage{amsfonts}
\usepackage{amsthm}
\usepackage{algorithm}
\usepackage{algpseudocode}
\usepackage{natbib}
\usepackage{hyperref}
\usepackage{cleveref}
\usepackage{multirow}
\usepackage{comment}
\usepackage{svg}
\usepackage{authblk}

\title{Causal Responder Detection}

\newcommand{\bb}[1]{\mathbb{#1}}
\newcommand{\ind}{\mathrm{\mathbf{1}}}

\newtheorem{lemma}{Lemma}

\newtheorem{remark}{Remark}

\author[1]{Tzviel Frostig\thanks{tzviel@phasevtrials.com}}
\author[1]{Oshri Machluf\thanks{oshri@phasevtrials.com}}
\author[1]{Amitay Kamber}
\author[1]{Elad Berkman}
\author[1]{Raviv Pryluk}

\affil[1]{Research Department, PhaseV}

\begin{document}

\maketitle

\begin{abstract}
We introduce the causal responders detection (CARD), a novel method for responder analysis that identifies treated subjects who significantly respond to a treatment. Leveraging recent advances in conformal prediction, CARD employs machine learning techniques to accurately identify responders while controlling the false discovery rate in finite sample sizes. Additionally, we incorporate a propensity score adjustment to mitigate bias arising from non-random treatment allocation, enhancing the robustness of our method in observational settings. Simulation studies demonstrate that CARD effectively detects responders with high power in diverse scenarios.
\end{abstract}

\section{Introduction} \label{sect:intro}

Personalized medicine is expected to advance healthcare in the near future \citep{vicente2020personalised}.
In contrast to a one-size-fits-all approach, personalized medicine advocates for treatments tailored to individual patients based on their clinical characteristics. 

Responder analysis in clinical trials is a method used to evaluate the effectiveness of a treatment by identifying and analyzing the subset of participants who respond significantly to the treatment \citep{henschke2014responder}. This approach contrasts with the traditional method of evaluating average effects across all participants, which can sometimes obscure the benefits seen in a particular group of responders \citep{guyatt1998interpreting}. 
This approach is particularly important in trials with heterogeneous treatment effects, where understanding individual responses can lead to more effective therapies. Responder analysis is a common practice in clinical trials \citep{moore2010responder, straube2010pregabalinResponder, chuang2022dupilumab}. In the context of this manuscript we focus on minimal evidence for responsiveness rather than clinically meaningful responses \citep{snapinn2007responder}. 

Another approach to understanding heterogeneous treatment effects is estimating the conditional average treatment effect (CATE), which focuses on understanding how the effect of a treatment varies across different segments of the population, rather than only calculating the average treatment effect (ATE). Unlike ATE, which provides a single overall effect, CATE enables the identification of treatment benefits that may be specific to subgroups defined by their unique characteristics \citep{angus2021heterogeneity}.

CATE and responder analysis complement each other. CATE offers insights into the heterogeneous treatment effects in terms of the expected values of certain population segments. Meanwhile, responder analysis provides additional context by indicating the proportion of total responders and extends beyond the expected value. For instance, consider a scenario where a certain treatment increases the variability within a subgroup of the population without necessarily affecting their expected value. Some subjects in this subgroup would be identified as responders, however the CATE estimation will offer no information regarding it. 

Individual treatment effects (ITE) provide a comprehensive view of the treatment response. \cite{lei2021conformalITE} proposed a method for constructing confidence intervals for the ITE distribution using split conformal prediction and counterfactual prediction \citep{papadopoulos2008inductive}. These confidence intervals offer more information than the CATE point estimate alone. However, they often lack the statistical power to detect differences between treated and untreated responses. These confidence intervals can be used for responder analysis, where subjects with confidence intervals that exclude $0$ can be determined as responders. 


CARD (causal responder detection), our proposed responder analysis method, builds on AdaDetect (adaptive detection, 
\cite{marandon2024adaptiveAdadetect}). AdaDetect is a conformal prediction method for detecting out-of-distribution samples using an adaptive non-conformity score.
CARD offers two substantial innovations over AdaDetect for the task of responder identification. First, we suggest a specialized scorer aimed at discovering responders. This scorer uses recursive partitioning of the feature space to maximize the differences in response between treated and untreated subjects within the nodes, similar to the causal tree algorithm \citep{athey2016recursiveCT}. This approach increases the power substantially compared to AdaDetect, which utilizes off-the-shelf classifiers as scorers. Second, we address the issue of potential bias from non-random allocation of treatment by applying a propensity score based adjustment \citep{tibshirani2019conformalweighted}.

The rest of \cref{sect:intro} provides background on the inference task at hand. In \cref{sect:related_work} we overview AdaDetect and causal trees, as they serve as building blocks for our method. In \cref{sect:suggested_method} we present CARD, discuss its components, suggest an adjustment to handle cases where the treatment assignment depends on the covariates. In \cref{sect:simulations} we examine the performance of CARD in RCT and observational study settings.  Finally, we conclude with a discussion in \cref{sect:discussion}.

\subsection{Causal Inference}

In this paper, we follow the potential outcomes framework \citep{imbens2015causal} with a binary treatment. Given $n$ subjects, we denote by $T_i \in {0, 1}$ the binary treatment indicator, by $(Y_i(1), Y_i(0))$ the pair of potential outcomes, and by $X_i$ the vector of other covariates of length $p$. We assume that
\begin{equation*}
(Y_i(1), Y_i(0), T_i, X_i) \sim (Y(1), Y(0), T, X),
\end{equation*}

where $(Y(1), Y(0), T, X)$ denotes a generic random vector. Our focus will be on a binary treatment, $T \sim \text{Ber}(e(x))$, where $e(x)$ is the propensity score. Under the commonly assumed stable unit treatment value assumption (SUTVA), the observed data set comprises triplets $(Y_{\text{obs}}, T_i, X_i)$ where $Y_{\text{obs}} = T_i \times Y_i(1) + (1 - T_i) \times Y_i(0)$.

The ITE $\tau_i$ is defined as
\begin{equation*}
\tau_i \equiv Y_i(1) - Y_i(0).
\end{equation*}

By definition, for each unit, only one potential outcome is observed while the other remains unobserved. Consequently, the ITEs are unobserved and must be inferred. We also assume strong ignorability throughout the paper:
\begin{equation*}
(Y(1), Y(0)) \perp T \mid X.
\end{equation*}

Strong ignorability assumes that there are no unmeasured confounders affecting both the treatment assignment and the potential outcomes. Under this assumption, the treatment assignment is essentially randomized conditional on the covariate values. These assumptions allow us to claim that
$$Y_i(1)|X=x \sim Y_i|X_i = x, T_i = 1, \quad Y_i(0)|X=x \sim Y_i|X_i = x, T_i = 0,$$ making it possible to identify moments of interest in the distribution of $Y(1) - Y(0)$. Typically, the interest lies in estimating the CATE,
\begin{equation*}
    \tau(x) = \mathbb{E}(Y(1) - Y(0) | X=x) = \mathbb{E}(Y| X = x, T = 1) - \mathbb{E}(Y | X = x, T = 0),
\end{equation*} where the second equality is due to the strong ignorability assumption. The estimation of CATE garnered much attention in recent years, multiple methods have been developed to estimate CATE using machine learning \citep{kunzel2019metalearners,  wager2018estimationCF, nie2021quasiRlearner}. For example, the S-learner consists of training an outcome prediction algorithm $\hat{\mu}(X, T)$, which aims to predict $E(Y|X = x, T = t)$. The CATE estimate is then given by
$$\hat{\tau}(x) = \hat{\mu}(x, T = 1) - \hat{\mu}(x, T = 0).  $$ Other parameters of the ITE distribution (besides CATE) are also valuable targets for estimation. \cite{fort2016unconditional} proposed estimating the conditional quantile treatment effect. 

\subsection{Inferential Goal}

The objective of responder analysis is to determine whether each treated observation is a responder, as defined by:

\begin{equation*} 
H_{0, i}: Y_i(1) \sim Y_i(0).
\end{equation*} Under the strong ignorability assumption, the above is equivalent to 
\begin{equation}\label{eq:hypo_test}
H_{0, i}: (Y_i(1),X_i) \sim P_{Y| X , T = 0} \times P_{X|T=1}.
\end{equation} One could argue that the objective should be to test $Y_i(1) \sim P_{Y|X=X_i, T = 0}$. However, such conditional testing requires assumptions of smoothness and increasingly large sample sizes \citep{foygel2021limits}. Since the tested distribution also depends on $P_{X|T=1}$, in some settings, testing individual hypotheses becomes meaningless (see \cref{appendix:example}). Consequently, we aim to control the false discovery rate (FDR) across all treated observations.

\begin{remark}
    The hypothesis presented in \cref{eq:hypo_test}, is closely related to the task of testing if the ITE is $0$, but does not cover all possible scenarios. For example, suppose that $Y_i(1),Y_i(0) \sim N(0,1) \; i = 1,2$ are independent. It is clear the ITE is not zero, yet the hypothesis remains true. 
\end{remark}


\section{Related Work} \label{sect:related_work}

In the following section we provide the relevant background for CARD. 

\subsection{Causal Tree and Forest}

Causal tree (CT) is a type of decision tree specifically design for estimating CATE \cite{athey2016recursiveCT}. A tree is a partition, $\Pi$, of the covariate space $X$. We denote by $l(x; \Pi)$ the partition $x \in l, l \in \Pi$. A partition is obtained by applying a partitioning algorithm, which is used to minimize a loss function on some sample $(y_i,x_i,t_i), i\in \mathcal{I}$. Denote by $\mathcal{I}_j = \{i: t_i = j\}, \; j = 0,1$, and $n_j = |\mathcal{I}_j|$. Finally, denote $L(\mathcal{I}_j, x, \Pi) = \{i: i \in \mathcal{I}_j, x_i \in l(x; \Pi) \}$.
In CTs at each partition the ATE is estimated by,
\begin{equation}  \label{eq:prediction_ct}
\hat{\tau}(x_i; \Pi) =  \frac{1}{| L(\mathcal{I}_0, x_i, \Pi)|} \sum_{j \in L(\mathcal{I}_0, x_i, \Pi) } y_j - \frac{1}{| L(\mathcal{I}_1, x_i, \Pi)|} \sum_{v \in L(\mathcal{I}_1, x_i, \Pi)} y_v. 
\end{equation} The partition, $\Pi$ is obtained by minimizing the modified CATE mean square error loss, 
\begin{equation} \label{eq:loss_ct}
\begin{split}
     Loss(\Pi, \mathcal{I} )  = & -\frac{1}{n} \sum_{i \in \mathcal{I}_0 \cup \mathcal{I}_1} \hat{\tau}^2(x_i; \Pi) \\ & + \frac{2}{n}  \sum_{l \in \Pi} \bigg( \frac{ n_0\mathrm{\hat{V}}( \{y_i: i \in \mathcal{I}_0, x_i \in l \})}{n}  \\ & +  \frac{n_1 \mathrm{\hat{V}}( \{y_i: i \in \mathcal{I}_1, x_i \in l \}) }{n} \bigg),
\end{split}
\end{equation}
where $n=n_0+n_1$ is the total number of observations, and $\hat{V}$ is the standard sample variance estimator. To tackle the selective inference issue, the authors suggest splitting the data into a training and test sets. The training set is used to find the partitions that minimize \cref{eq:loss_ct}, and the test-set is used for estimating the CATE within each terminal node (\cref{eq:prediction_ct}). 

The use of sample splitting ensures that the estimates of the CATE in the terminal nodes are unbiased. Thus, any standard inference approach can be used to test the difference in means between the treated and untreated subjects in each terminal leaf using the test set. The trade-off is that there are fewer observations for finding the optimal partition. Furthermore, the method is effective only when generating a single tree. When growing a forest, sample-splitting is not sufficient to ensure valid testing, and different methods for estimating the variance must be applied. These methods have only asymptotic guarantees \citep{wager2018estimationCF}.

\subsection{Conformal Prediction}

Conformal prediction is a framework that provides a method to generate a prediction set that contains the true target value with a $1 - \alpha$ confidence \citep{angelopoulos2023conformal}. We focus on inductive conformal prediction in which the calibration set is independent of the samples used for training $\hat{f}$ \citep{papadopoulos2008inductive}. Suppose there is a predictive model $\hat{f}$, trained on some $(X_i, Y_i)$, where $Y_i$ is a continuous response and $X_i$ are the covariates. Conformal prediction requires a specification of a non-conformity score, $s(x, y)$, (where larger values correspond to worse agreement between $x$ and $y$) and a calibration set $(X_i, Y_i), \; i = 1, \ldots, k$. The non-conformity score is often learnt based on a training set. To test a new observation, the non-conformity score is computed for the calibration set, $s(\hat{f}(X_1), Y_1), \ldots, s(\hat{f}(X_k), Y_k)$. Assuming that a new observation is from the same distribution as the calibration set, then their non-conformity scores are also distributed identically, thus, \begin{equation*} \label{eq:conformal_quantile}
\mathbb{P}\left(s(X_{test}, Y_{test}) > q\left((1-\alpha)\frac{k + 1}{k} \right) \right) \leq \alpha, 
\end{equation*} where $q(p)$ is the empirical quantile based on the calibration set non-conformity scores. This can be used to test $H_0: Y_{test}|X_{test} \sim  Y_i | X_i$, or by inverting the acceptance region generating prediction-intervals  \citep{shafer2008tutorial, haroush2021statistical}. 
The choice of the score function is critical for the performance of the conformal predictor, \cite{romano2019conformalized} suggested conformal quantile regression (CQR), an adaptive score function based on quantile regression. This allowed for the conformal interval to vary as function of $x$, making it adaptive when facing heteroscedasticity. Later, \cite{romano2019conformalized} generalized the approach and suggested conformal-intervals that are based on the conditional histogram of $Y|X$. 

To test $H_0: Y_{test} \sim  P_{Y|X} \times P_{X}$, using the procedure described above, one must assume that $X_{test} \sim P_{X}$. When one is unwilling to make such assumption, weighted conformal prediction can be used. \cite{tibshirani2019conformalweighted} suggested a modification to conformal prediction when facing covariate shift, where the score is weighted according to the likelihood ratio $w(x) =  \frac{\mathrm{d} {P}_{test}(x)}{\mathrm{d} {P}_0(x)}$, where ${P}_0$ is the distribution of the calibration set covariates.

\cite{lei2021conformalITE} proposed confidence intervals for the ITE. In the process, they demonstrated how to construct counterfactual confidence intervals. The method builds on the CQR and utilizes the weighted conformal prediction to handle propensity. 

\subsubsection{AdaDetect}

Suppose there are two samples, a reference sample $\mathcal{I}_0 = \{i: X_i \sim {P}_0 \}$, and a test sample $\mathcal{I}_1 = \{i: X_i \sim {P}_i\}$, where ${P}_i$ can either be ${P}_0$ or some other distribution.  
The interest lies in identifying the specific observations in the test sample who differ from the reference sample, i.e., testing, 
\begin{equation} \label{eq:ood_test}
    H_{0,i}: X_i \sim P_0, \forall i \in \mathcal{I}_1.
\end{equation}

Conformal prediction can be used for this purpose. First, a non-conformity score, $s$, which measures the degree of disagreement of an observation being sampled from $P_0$ is applied to $\mathcal{I}_0$ to estimate its empirical distribution under the null hypothesis \citep{haroush2021statistical}. 
Then, the same non conformity is applied to each observation in $\mathcal{I}_1$. A p-value for each observation obtained according to $\mathrm{pv}_i = \frac{1 + \sum_{j \in \mathcal{I}_0}  \ind\left( s(x_j) > s(x_i) \right)}{|\mathcal{I}_0| + 1},$ where $\ind$ is the indicator function. 

\cite{bates2023testing} have shown that the resulting p-values adhere to a specific type of dependency and that the Benjamini-Hochberg (BH) multiplicity adjustment \citep{benjamini1995controlling} can be safely applied to control the FDR. While the procedure is valid, it still remains to decide on $s$, as different scores will be powerful for different ${P}_1$ and ${P}_0$.

Recently, \cite{marandon2024adaptiveAdadetect} proposed AdaDetect, an adaptive method for learning the non-conformity score $s$ for such hypotheses. The non-conformity score is learned using a classifier aimed at discriminating between the reference and test sets. However, even when the null hypothesis of \cref{eq:ood_test} is true for all observations in the test set, the application of this classification method may still result in different distribution of non-conformity scores between the two samples. 

To address this issue, AdaDetect employs knockoffs, observations designed to be indistinguishable from actual test observations to the classifier, yet for which the null hypothesis holds. These knockoffs generate non-conformity scores that are distributed similarly to the scores of true test observations under the null hypothesis, thus enabling more accurate hypothesis testing. To overcome the issue, the method employs knockoffs, observations that appear to the classifier as if they belong to $\mathcal{I}_1$ but for which \cref{eq:ood_test} holds true. These knockoffs generate non-conformity scores that are distributed similarly to the scores of true test observations under the null hypothesis, thus enabling more accurate hypothesis testing.

The procedure involves sampling $k$ observations from $\mathcal{I}_0$ (denote as $\mathcal{I}_{ko}$), that will serve as our knockoffs, we will combine them with $\mathcal{I}_1$, such we are left with $\mathcal{I}_1^* = \mathcal{I}_{ko} \cup \mathcal{I}_1$ and $\mathcal{I}_0^* = \mathcal{I}_0 / \mathcal{I}_{ko}$. 

The non-conformity score is learnt by training a classifier aimed at discriminating between $\mathcal{I}_0^*$ and $\mathcal{I}_{1}^*$, where higher values indicate the sample is more likely to be from $\mathcal{I}_{1}^*$. The prediction will also serve as our non-conformity score. Intuitively, $s(X_i), i \in \mathcal{I}_1 : X_i \sim {P}_0$, are exchangeable with $s(X_j), j \in \mathcal{I}_{ko}$ under $H_0$, thus a p-value can be obtained by, 

\begin{equation} \label{eq:unweighted_adadetect_pval}
    \mathrm{pv}_i = \frac{1 + \sum_{j=1}^k  \ind \left( s(x_j) > s(x_i) \right)}{k + 1}.
\end{equation}

The use of a classifier generates an adaptive non-conformity score, which was shown to asymptotically achieve the optimal Neyman-Pearson test power \citep{marandon2024adaptiveAdadetect}.

\section{CARD} \label{sect:suggested_method}

CARD applies a targeted approach using a specialized scorer to improve the power in detecting changes in response distribution for treated subjects. Similar to AdaDetect, it employs a classifier to detect the observations of interest and knockoffs to obtain FDR control in finite samples. CARD employs 
scorer tailored to the specific structure of the problem, substantially improving its power over AdaDetect which uses generic classifiers. 

We initially discuss the suggested method in an RCT setting, which naturally avoids confounding due to differences in the distributions of $X|T = 1$ and $X|T = 0$. In \cref{sect:prop_adj}, we also discuss approaches for addressing such differences in non-randomized settings.

We define $\mathcal{I}_0$ and $\mathcal{I}_1$ as the indices for untreated and treated observations, respectively. Additionally, define the set of treated observations indices where (\cref{eq:hypo_test}) holds true,   $\mathcal{H}_0 = \{i\in \mathcal{I}_1: Y_i|T_i=1, X_i \sim Y_i|T_i=0, X_i\}$.

\subsection{Inference in RCTs} 

To test the hypotheses \cref{eq:hypo_test}, the untreated subjects are split into two groups, $\mathcal{I}_0 = \mathcal{I}_{ko} \cup \mathcal{I}_0^*$. A classifier is trained to differentiate between $\mathcal{I}^*_1=\mathcal{I}_1\cup\mathcal{I}_{ko}$ and $\mathcal{I}_0^*$. Let $s(x_i, y_i)$ be the resulting score according to the classifier. Finally, p-values are obtained as in \cref{eq:unweighted_adadetect_pval}, 

\begin{equation*} 
    \mathrm{pv}_i = \frac{1 + \sum_{j \in \mathcal{I}_{ko}}  \ind \left( s(x_j, y_j) > s(x_i, y_i) \right)}{k + 1}.
\end{equation*}

In RCTs, randomization ensures that $X|T=1 \sim X|T=0$. If the null hypothesis (\cref{eq:hypo_test}) is true for observation $i$, then $Y_i| T_i = 1, X_i \sim Y_i| T_i = 0, X_i$. Thus, $ \{ (X_i, Y_i)\}_{i \in \mathcal{I}_{ko}}$ are exchangeable with $ \{ (X_j, Y_j)\}_{j \in \mathcal{I}_{1} \cap \mathcal{H}_0}$. If the classifier is invariant (as defined by eq. 8 in \citep{marandon2024adaptiveAdadetect}), a property of most practical classifiers, then according to Lemma 3.2 and Theorem 3.3 of \citep{marandon2024adaptiveAdadetect}, the BH procedure applied to the resulting p-values will control the FDR. 

\begin{figure}[ht]
    \centering
    \includegraphics[width=0.95\linewidth, height=3.5in ]{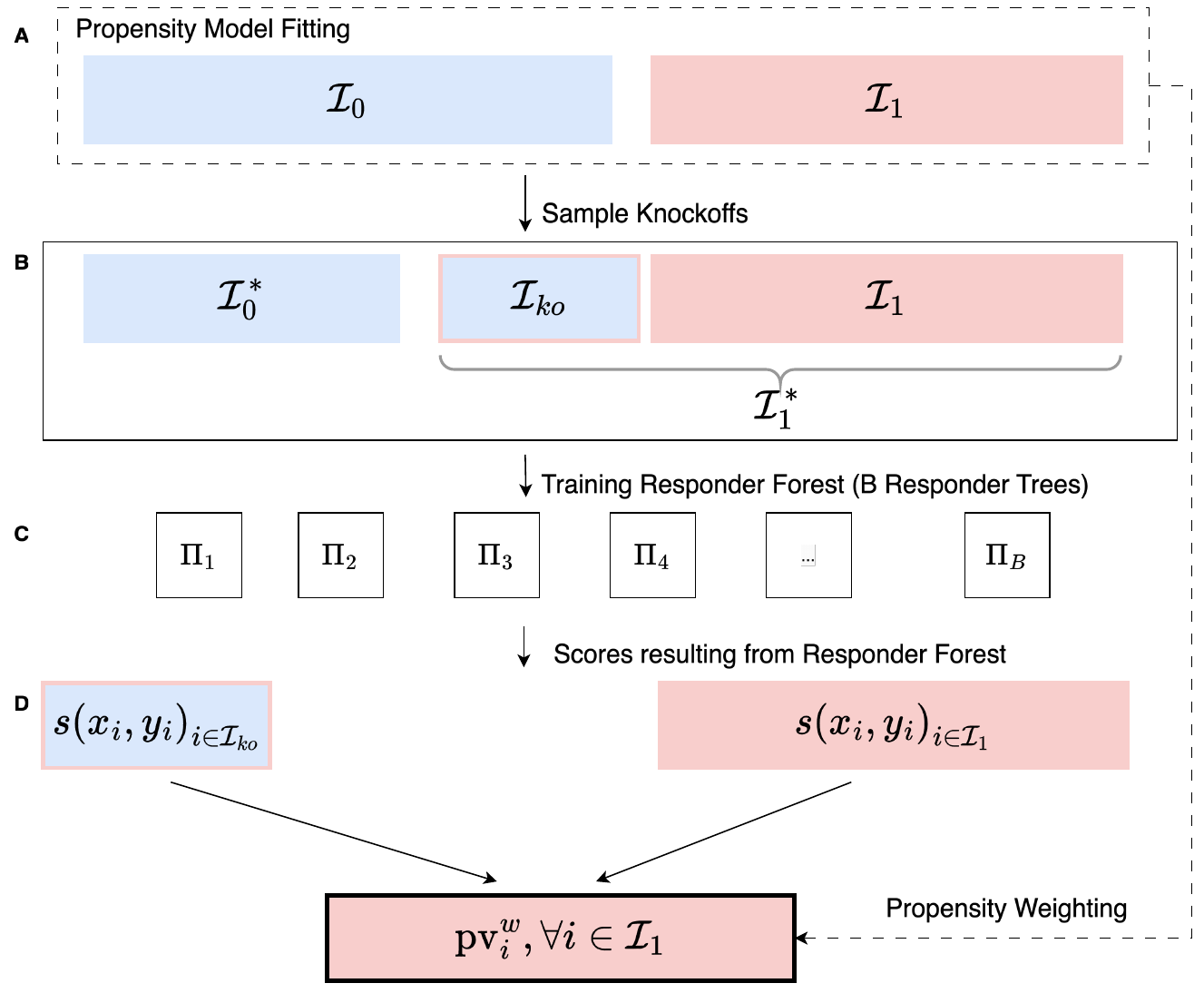}
    \caption{An illustration of CARD is provided where red fill indicates treated samples, and blue fill represents untreated samples. Knockoffs are depicted with a red frame. In step A, the propensity model is fit, and knockoffs are sampled from the untreated observations. Step B involves preparing the data for the classifier by combining the knockoffs with the treated observations. In step C, the responder forest is trained to classify between $\mathcal{I}^*_0$ and $\mathcal{I}^*_1$. Finally, in step D, the predictions from the responder forest are used as non-conformity scores. These scores are then weighted according to the propensity score to calculate p-values.}
    \label{fig:conditional_dists}
\end{figure}

To handle the setting of observational studies, where $X|T=1 \not\sim X|T=0$, we suggest a weighted version of Adadetect based on weighted conformal prediction \citep{tibshirani2019conformalweighted}, see \cref{sect:prop_adj}. A graphical illustration of the procedure is provided in \cref{fig:conditional_dists}.

\subsection{Scorer}

To increase the power of the AdaDetect, we suggest learning the non-conformity scores using a responder tree.  
The responder tree mirrors the principles of CTs, wherein the data is recursively partitioned on $X$ to minimize the overall $\mathrm{Loss}_y$ across partitions until certain stopping criteria, such as reaching a maximum tree depth or convergence of loss, are met. While the splits are done according to $X$ the evaluation in the terminal leaves is based solely on $Y$ and $T$. 

The splits the responder tree makes are driven by the loss on the terminal nodes (where only $Y$ is considered). The aim of the tree is to best classify observations by their treatment assignment (treated or not). The loss the responder tree attempts to minimize is
\begin{equation} \label{eq:loss_responder_tree}
     \mathrm{Loss}(\Pi, \mathcal{I}_0, \mathcal{I}_1 ) =  \sum_{l \in \Pi} c_l  \mathrm{Loss}_y( \Pi_y, \{y_i: x_i\in l, i\in \mathcal{I}_0\}, \{y_i: x_i \in l, i\in \mathcal{I}_1\}), 
\end{equation} where $\Pi_y$ is a partition that spans $Y$,  $\mathrm{Loss}_y$ can be any loss applicable for classification such as Gini or entropy loss, and $$c_l = \frac{|\{i\in \mathcal{I}_0 \cup \mathcal{I}_1:x_i\in l\}|}{|\mathcal{I}_0 \cup \mathcal{I}_1|}. $$

We minimize the specified loss (\cref{eq:loss_responder_tree}) using recursive partitioning, a standard approach to optimize classification trees \citep{rokach2005top}. The primary partitions are based on 
$X$. For each subset of observations defined by these partitions, we fit a shallow classification tree (depth of 2) where $Y$ is used to classify $T$. The loss associated with each partition is calculated based on the weighted classification loss of the shallow trees applied to each subset. Partitioning continues provided there is an improvement in the loss or until predefined stopping criteria are met, such as a meeting the minimal number of observations in the terminal nodes or reaching the maximal tree depth.

While a single responder tree is described, we can also consider a forest of scorers, where in a similar vain to Random Forest (RF), a bootstrap sample of observations and a sample of features are taken to grow each tree \citep{hastie2009random}. The score is the average prediction across the different trees as standard in classification random-forest.

\subsection{Propensity Adjusted CARD} \label{sect:prop_adj}

If the propensity is a function of $X$ and $Y(0)\not \perp X$ then the knockoffs are no longer exchangeable with the treated observations under the null. To overcome this issue, we suggest a propensity adjusted p-value, 


\begin{equation} \label{eq:weighted_pvalues}
    \mathrm{pv}^w_i = w_i^* + \sum_{j\in \mathcal{I}_{ko}}  w_j^* \cdot \ind\left(s(x_j, y_j) > s(x_i, y_i)\right),
\end{equation}
where the weights are defined by $w_j = \frac{e(x_j)}{1 - e(x_j)}$ $j\in \{i\}\cup \mathcal{I}_{ko}$ and $w_j^* = \frac{w_j}{\sum_{l\in \{i\}\cup \mathcal{I}_{ko}} w_l}$. This is similar to the weighting used in \citep{lei2021conformalITE}.

\begin{lemma} \label{lemma:prop_adj}
    Under the null hypothesis (\cref{eq:hypo_test}), the weighted p-values (\cref{eq:weighted_pvalues}) control the type I error at level $\alpha$, 

    $$ \bb{P}(\mathrm{w-pv}_i < \alpha) \leq \alpha .$$
\end{lemma} The proof is given in \cref{appendix:proof}. It should be noted, that the resulting p-values have no theoretical control of the FDR in finite sample, however, in practice they indeed control the FDR. Furthermore, it is shown that as $n \rightarrow \infty$, the BH procedure using the weighted p-values recover the theoretical control of the FDR \citep{jin2023model}.

Although results are typically presented with oracle propensity, in practice, the propensity score is estimated from the same data we intend to analyze. To reduce over-fitting, we fit the propensity estimator using $m$-fold fitting \citep{jacob2020cross}, training on $m-1$ folds and estimating the propensity on the remaining fold. We examine the empirical validity of this adjustment in \cref{sect:prop_adj_sim}.

\section{Simulations} \label{sect:simulations}

\subsection{RCT Simulations} \label{sect:rct_sim}

We begin by investigating the performance of the suggested method in RCT settings using a simulation study. We use a variant of the data generation process that was used by \cite{lei2021conformalITE} and \cite{wager2018estimationCF}.
The data generation process is outlined as follows: We sample $X^* \sim N(0, \Sigma_{p\times p})$, where $\Sigma_{i,j} = \rho^{|i - j|}$. To obtain $X$ we transform $X_i = \Phi(X^*_i)$. The potential outcomes are generated according to, 
$$Y(0) = 4\cdot(X_1 + X_2) + \epsilon, \quad Y(1) = 4\cdot(X_1 + X_2) + r f(X_1)\cdot f(X_2) + \epsilon,$$ where $\epsilon \sim N(0, \sigma(x)^2 )$ and $$f(x) = \ind (x > 0.5) \cdot \frac{2}{1 + \exp(-12 \cdot (x - 0.5))}.$$ 

Treatment assignments are randomly determined following a Bernoulli distribution, $\operatorname{Ber}(0.5)$. We explore variations in the parameters: $r \in \{-1, +1\}$ to assess the effect of signal sign, $\sigma(x) \in \{1, -2 \cdot \ln(x)\}$ to compare homoscedastic and heteroscedastic settings, $\rho \in \{0, 0.9\}$ to distinguish between independent and dependent settings, and $p \in \{10, 100\}$ to evaluate the impact of the number of features. We vary the number of observations, $n \in \{250, 500, 1000, 2000 \}$. Each setting is repeated $100$ times, and the averages are reported. 

The null hypothesis tested by the suggested method \cref{eq:hypo_test}, is false for $X_1 > 0.5, X_2 > 0.5$. To handle multiplicity, we adjust the p-values using the BH procedure. The FDR is expected to controlled at level $0.1$. We compare CARD (using 20\% of the untreated subjects as knockoffs) with several other methods of detecting responders. 

\begin{enumerate}
    \item Global Testing - Testing if $Y_i, T_i = 1 \sim Y(0)$, which is also valid for testing \cref{eq:hypo_test}. The p-value is obtained on the basis of the $Y(0)$ eCDF (see \cref{appendix:global}).
    \item AdaDetect \citep{marandon2024adaptiveAdadetect} - Applying the original AdaDetect method with RF as the classifier. 20\% of the untreated observations are used as knockoffs. 
    \item CQR \citep{romano2019conformalized} - We apply the CQR non-conformity score to obtain p-values (see \cref{section:cqr}), this is equivalent to counter factual confidence-intervals suggested by \cite{lei2021conformalITE}. Boosting is used for the estimation of the conditional quantiles, 80\% of the untreated sample is used for estimation and 20\% for testing. 
 \end{enumerate}

All methods are valid for testing \cref{eq:hypo_test} and controling FDR in finite samples. Other methods of CATE estimation can also be used for testing for responders such as CF, BART and bootstrap with Meta-Learners \citep{wager2018estimationCF, chipman2012bart, kunzel2019metalearners}. However, \cite{lei2021conformalITE} showed they lack finite coverage control, and thus are omitted from this and subsequent simulation studies.

The FDR and additional power results are given in \cref{appendix:rct_sim}. The use of AdaDetect with RF as the classifier yields poor results, which is unsurprising as RF attempts to detect differences across all features, unlike competing methods focusing on $Y$ or $Y|X$.

Responder knockoffs is powerful in all scenarios, significantly outperforming competing methods in most cases, except in homoscedastic scenarios with positive signals where they are surpassed by the CQR and global methods for $n < 2000$. It's power increases the most as the sample size increases. 

As implied by its name, the global method performs well when responders are discernible solely by considering $Y$, particularly in scenarios with positive signal sign and homoscedastic noise. Because the global method does not split the data, it significantly outperforms other methods in this specific scenario.
However, when the differences in distributions are only discernible upon conditioning on $X$ (see \cref{appendix:global} for an illustrative example), the method has virtually no power. 

Unlike CARD, the CQR non-conformity score is learnt only on the basis of the untreated distribution. This leads to much better performance of CARD across all settings except for positive signal and homoscedastic noise.

\begin{figure}[ht]
    \centering
    \includegraphics[width=0.95\linewidth ]{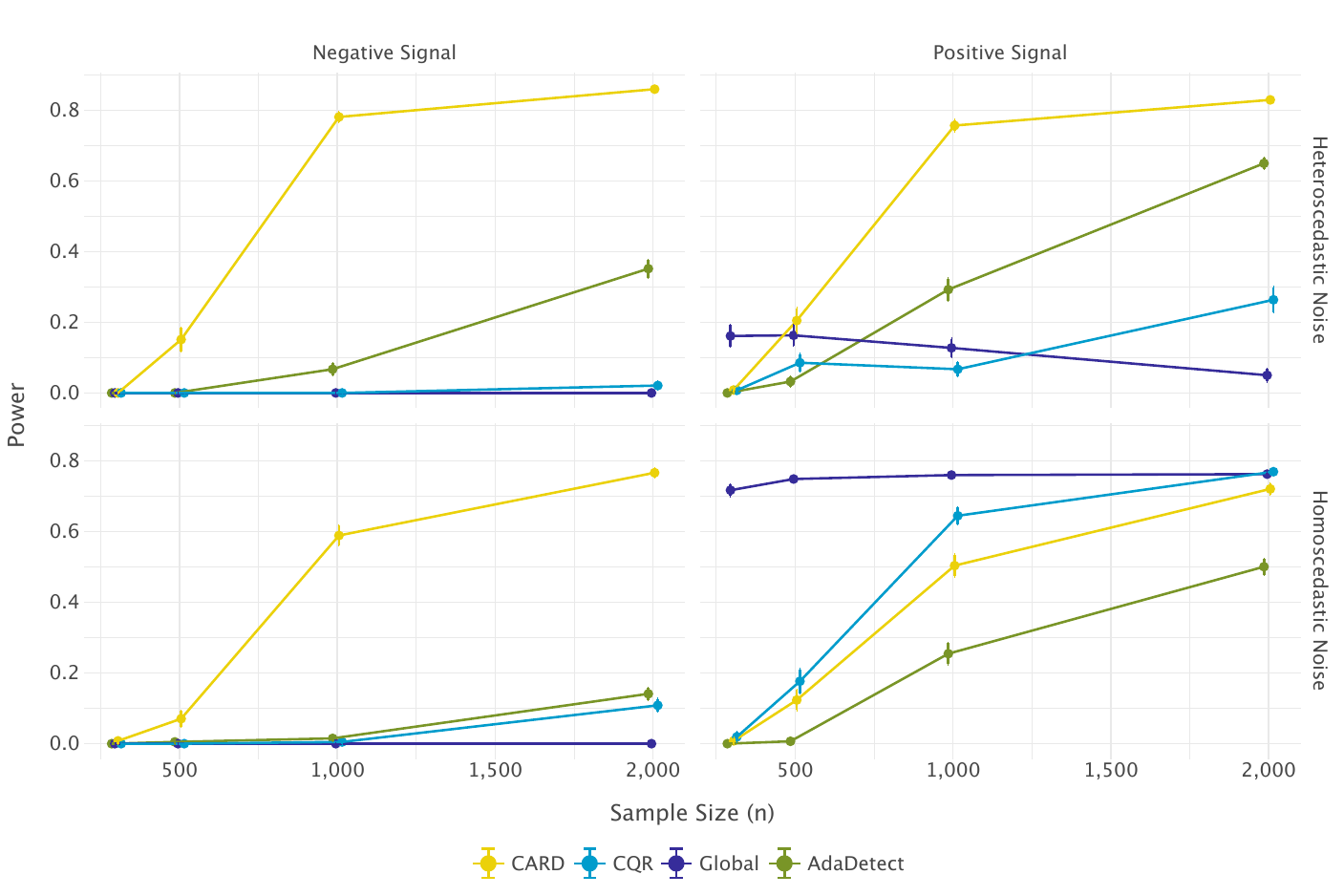}
    \caption{Power analysis of various methods for different sample sizes. The plots display the power as a function of sample size (n) in the $p=10$ scenario. The power of CARD is higher in all scenarios except for the homoscedastic noise and positive signal, where it reaches parity when $n$ is sufficiently large. Intervals indicate $\pm1$ SD around the estimate. }
    \label{fig:rct_simulation_plot}
\end{figure}

\subsection{Observational Study Simulation} \label{sect:prop_adj_sim}

Although the asymptotic control of the FDR in the weighted CARD procedure is theoretically guaranteed, the effects of weighting in finite samples and in practical settings remains unknown.

We employ the same data generation protocol as outlined in \cref{sect:simulations}; changing only the propensity function to mimic an observational study with confounders, $e = 
\frac{1}{1 + \exp(1.75 \cdot X_1 - 0.825)}$. The propensity function ensures that the probability of receiving treatment is $0.5$. We present results for the $p = 10$ and positive signal scenario, as the results are consistent across scenarios. Furthermore, we restrict our simulations to CARD.

We consider four methods of handling non-trivial propensity: 1) None, ignoring propensity and applying CARD without any adjustment; 2) Oracle adjustment, where weighting is based on true propensity; 3) RF classifier propensity adjustment; 4) Logistic Regression propensity adjustment. The propensity models are fit using 10-fold cross-fitting.

\begin{figure}[ht]
    \centering
    \includegraphics[width=0.75\linewidth ]{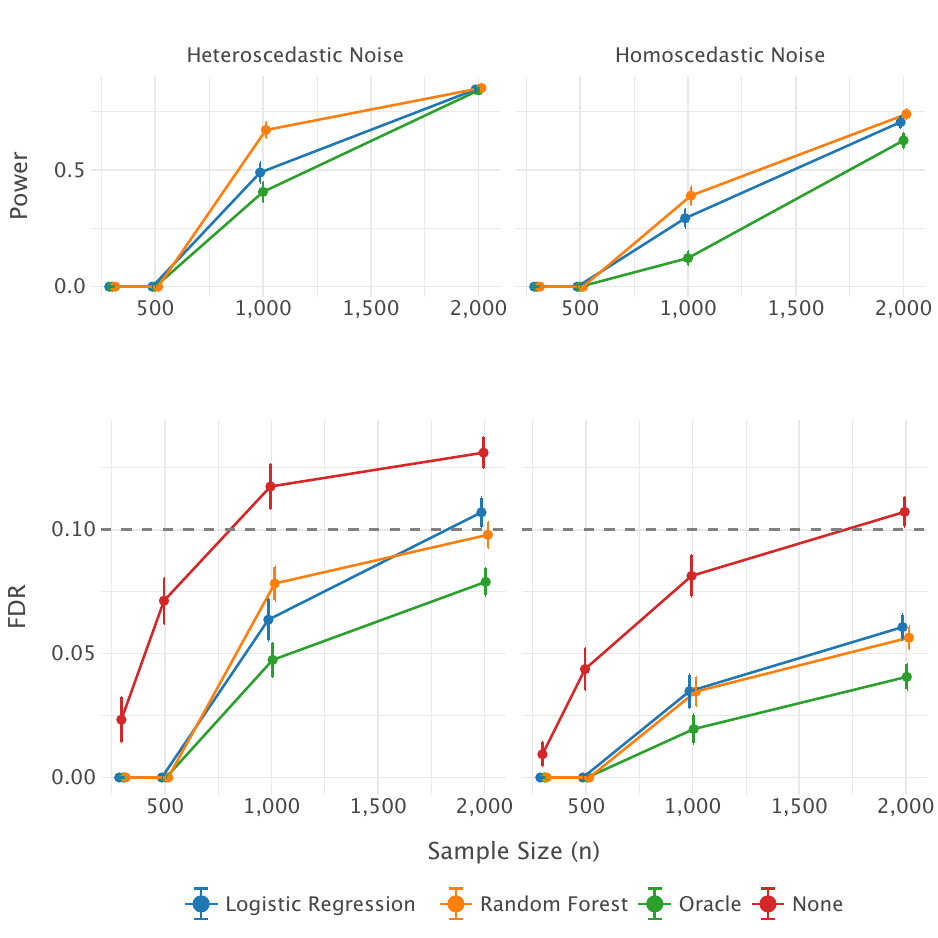}
    \caption{Power and FDR of the various methods of handling propensity. The None method is removed from the power (upper row) plots, since it fails to control the FDR at the expected level (bottom row). Intervals indicate $\pm1$ SD around the estimate. }
    \label{fig:obs_simulation}
\end{figure}

The results, shown in \cref{fig:obs_simulation}, indicate that all methods except 'None' control the FDR at the expected level of $0.1$. The estimating methods exhibit higher power than the oracle, a result attributed to cross-fitting and errors in propensity estimation. However, since the method is generally conservative, it controls the FDR at the expected level, which is promising for real-world applications.

\section{Discussion} \label{sect:discussion}

CARD is a powerful method for identifying responders within a treatment group. It offers FDR control in finite samples and is applicable in both observational studies and RCTs. The detection of responders provides a perspective that complements the analysis of CATE. While CATE measures the average effect of treatment for each segment of the population, CARD estimates the proportion of affected samples. For example, consider a clinical trial where for a particular subgroup in the population there is substantial CATE; however, responder analysis might reveal that only a small fraction of this subgroup truly benefits from the clinical treatment.
Moreover, the inference on CATE is restricted to asymptotic regimes (where the sample size increases to infinity), while CARD provides finite sample control of the FDR. 

The main drawback of CARD is that it offers only marginal control over the type I error, and not conditional on the covariates, $X$. This results from the finite sample control provided by the method \citep{foygel2021limits}, a limitation also shared by counter-factual ITE confidence intervals \citep{lei2021conformalITE}. Extending the method to test conditional hypotheses remains an area for future research.

While CARD is presented as a method for detecting responders, it can also be applied to other use-cases:
\begin{enumerate}
    \item A global null test for treatment effectiveness, allowing for quick screening of various datasets.
    \item Identifying subspaces of $X$ in which the treatment and non-treatment response distributions differ. 
    A single responder tree tries to find subspace of $X$ in where the distributions $Y(0)|X \in A$ and $Y(1)|X \in A$  are different. Testing the difference between the two distributions using only the knockoffs in the identified subspace, $\mathcal{I}_{ko} \cap A$ provides a valid test for distribution differences given $A$. 
    \item Enriching untreated samples in trials by selecting control samples that are sufficiently similar from previous trials. –
\end{enumerate}
These applications also present further research opportunities. 

Finally, CARD provides an estimate of the proportion of responders while controlling for the FDR. Pairing this method with various approaches to constructing valid confidence intervals for the true discovery proportion \citep{hemerik2018false, millstein2022fdrci} can provide even additional information on the proportion of responders, and is a promising research avenue.

\bibliographystyle{plainnat}
\bibliography{references.bib}

\begin{thebibliography}{32}
\providecommand{\natexlab}[1]{#1}
\providecommand{\url}[1]{\texttt{#1}}
\expandafter\ifx\csname urlstyle\endcsname\relax
  \providecommand{\doi}[1]{doi: #1}\else
  \providecommand{\doi}{doi: \begingroup \urlstyle{rm}\Url}\fi

\bibitem[Angelopoulos et~al.(2023)Angelopoulos, Bates, et~al.]{angelopoulos2023conformal}
Anastasios~N Angelopoulos, Stephen Bates, et~al.
\newblock Conformal prediction: A gentle introduction.
\newblock \emph{Foundations and Trends{\textregistered} in Machine Learning}, 16\penalty0 (4):\penalty0 494--591, 2023.

\bibitem[Angus and Chang(2021)]{angus2021heterogeneity}
Derek~C Angus and Chung-Chou~H Chang.
\newblock Heterogeneity of treatment effect: estimating how the effects of interventions vary across individuals.
\newblock \emph{Jama}, 326\penalty0 (22):\penalty0 2312--2313, 2021.

\bibitem[Athey and Imbens(2016)]{athey2016recursiveCT}
Susan Athey and Guido Imbens.
\newblock Recursive partitioning for heterogeneous causal effects.
\newblock \emph{Proceedings of the National Academy of Sciences}, 113\penalty0 (27):\penalty0 7353--7360, 2016.

\bibitem[Bates et~al.(2023)Bates, Cand{\`e}s, Lei, Romano, and Sesia]{bates2023testing}
Stephen Bates, Emmanuel Cand{\`e}s, Lihua Lei, Yaniv Romano, and Matteo Sesia.
\newblock Testing for outliers with conformal p-values.
\newblock \emph{The Annals of Statistics}, 51\penalty0 (1):\penalty0 149--178, 2023.

\bibitem[Benjamini and Hochberg(1995)]{benjamini1995controlling}
Yoav Benjamini and Yosef Hochberg.
\newblock Controlling the false discovery rate: a practical and powerful approach to multiple testing.
\newblock \emph{Journal of the Royal statistical society: series B (Methodological)}, 57\penalty0 (1):\penalty0 289--300, 1995.

\bibitem[Chipman et~al.(2012)Chipman, George, and McCulloch]{chipman2012bart}
Hugh~A Chipman, Edward~I George, and Robert~E McCulloch.
\newblock Bart: Bayesian additive regression trees.
\newblock \emph{Annals of Applied Statistics}, 6\penalty0 (1):\penalty0 266--298, 2012.

\bibitem[Chuang et~al.(2022)Chuang, Guillemin, Bachert, Lee, Hellings, Fokkens, Duverger, Fan, Daizadeh, Amin, et~al.]{chuang2022dupilumab}
Chien-Chia Chuang, Isabelle Guillemin, Claus Bachert, Stella~E Lee, Peter~W Hellings, Wytske~J Fokkens, Nicolas Duverger, Chunpeng Fan, Nadia Daizadeh, Nikhil Amin, et~al.
\newblock Dupilumab in crswnp: responder analysis using clinically meaningful efficacy outcome thresholds.
\newblock \emph{The Laryngoscope}, 132\penalty0 (2):\penalty0 259--264, 2022.

\bibitem[Fort(2016)]{fort2016unconditional}
Margherita Fort.
\newblock Unconditional and conditional quantile treatment effect: Identification strategies and interpretations.
\newblock \emph{Topics in theoretical and applied statistics}, pages 15--24, 2016.

\bibitem[Foygel~Barber et~al.(2021)Foygel~Barber, Candes, Ramdas, and Tibshirani]{foygel2021limits}
Rina Foygel~Barber, Emmanuel~J Candes, Aaditya Ramdas, and Ryan~J Tibshirani.
\newblock The limits of distribution-free conditional predictive inference.
\newblock \emph{Information and Inference: A Journal of the IMA}, 10\penalty0 (2):\penalty0 455--482, 2021.

\bibitem[Guyatt et~al.(1998)Guyatt, Juniper, Walter, Griffith, and Goldstein]{guyatt1998interpreting}
Gordon~H Guyatt, Elizabeth~F Juniper, Stephen~D Walter, Lauren~E Griffith, and Roger~S Goldstein.
\newblock Interpreting treatment effects in randomised trials.
\newblock \emph{Bmj}, 316\penalty0 (7132):\penalty0 690--693, 1998.

\bibitem[Haroush et~al.(2021)Haroush, Frostig, Heller, and Soudry]{haroush2021statistical}
Matan Haroush, Tzviel Frostig, Ruth Heller, and Daniel Soudry.
\newblock A statistical framework for efficient out of distribution detection in deep neural networks.
\newblock In \emph{International Conference on Learning Representations}, 2021.

\bibitem[Hastie et~al.(2009)Hastie, Tibshirani, Friedman, Hastie, Tibshirani, and Friedman]{hastie2009random}
Trevor Hastie, Robert Tibshirani, Jerome Friedman, Trevor Hastie, Robert Tibshirani, and Jerome Friedman.
\newblock Random forests.
\newblock \emph{The elements of statistical learning: Data mining, inference, and prediction}, pages 587--604, 2009.

\bibitem[Hemerik and Goeman(2018)]{hemerik2018false}
Jesse Hemerik and Jelle~J Goeman.
\newblock False discovery proportion estimation by permutations: confidence for significance analysis of microarrays.
\newblock \emph{Journal of the Royal Statistical Society Series B: Statistical Methodology}, 80\penalty0 (1):\penalty0 137--155, 2018.

\bibitem[Henschke et~al.(2014)Henschke, van Enst, Froud, and WG~Ostelo]{henschke2014responder}
Nicholas Henschke, Annefloor van Enst, Robert Froud, and Raymond WG~Ostelo.
\newblock Responder analyses in randomised controlled trials for chronic low back pain: an overview of currently used methods.
\newblock \emph{European Spine Journal}, 23:\penalty0 772--778, 2014.

\bibitem[Imbens and Rubin(2015)]{imbens2015causal}
Guido~W Imbens and Donald~B Rubin.
\newblock \emph{Causal inference in statistics, social, and biomedical sciences}.
\newblock Cambridge university press, 2015.

\bibitem[Jacob(2020)]{jacob2020cross}
Daniel Jacob.
\newblock Cross-fitting and averaging for machine learning estimation of heterogeneous treatment effects.
\newblock \emph{arXiv preprint arXiv:2007.02852}, 2020.

\bibitem[Jin and Cand{\`e}s(2023)]{jin2023model}
Ying Jin and Emmanuel~J Cand{\`e}s.
\newblock Model-free selective inference under covariate shift via weighted conformal p-values.
\newblock \emph{arXiv preprint arXiv:2307.09291}, 2023.

\bibitem[K{\"u}nzel et~al.(2019)K{\"u}nzel, Sekhon, Bickel, and Yu]{kunzel2019metalearners}
S{\"o}ren~R K{\"u}nzel, Jasjeet~S Sekhon, Peter~J Bickel, and Bin Yu.
\newblock Metalearners for estimating heterogeneous treatment effects using machine learning.
\newblock \emph{Proceedings of the national academy of sciences}, 116\penalty0 (10):\penalty0 4156--4165, 2019.

\bibitem[Lei and Cand{\`e}s(2021)]{lei2021conformalITE}
Lihua Lei and Emmanuel~J Cand{\`e}s.
\newblock Conformal inference of counterfactuals and individual treatment effects.
\newblock \emph{Journal of the Royal Statistical Society Series B: Statistical Methodology}, 83\penalty0 (5):\penalty0 911--938, 2021.

\bibitem[Marandon et~al.(2024)Marandon, Lei, Mary, and Roquain]{marandon2024adaptiveAdadetect}
Ariane Marandon, Lihua Lei, David Mary, and Etienne Roquain.
\newblock Adaptive novelty detection with false discovery rate guarantee.
\newblock \emph{The Annals of Statistics}, 52\penalty0 (1):\penalty0 157--183, 2024.

\bibitem[Millstein et~al.(2022)Millstein, Battaglin, Arai, Zhang, Jayachandran, Soni, Parikh, Mancao, and Lenz]{millstein2022fdrci}
Joshua Millstein, Francesca Battaglin, Hiroyuki Arai, Wu~Zhang, Priya Jayachandran, Shivani Soni, Aparna~R Parikh, Christoph Mancao, and Heinz-Josef Lenz.
\newblock fdrci: Fdr confidence interval selection and adjustment for large-scale hypothesis testing.
\newblock \emph{Bioinformatics Advances}, 2\penalty0 (1):\penalty0 vbac047, 2022.

\bibitem[Moore et~al.(2010)Moore, Moore, Derry, Peloso, Gammaitoni, and Wang]{moore2010responder}
R~Andrew Moore, Owen~A Moore, Sheena Derry, Paul~M Peloso, Arnold~R Gammaitoni, and Hongwei Wang.
\newblock Responder analysis for pain relief and numbers needed to treat in a meta-analysis of etoricoxib osteoarthritis trials: bridging a gap between clinical trials and clinical practice.
\newblock \emph{Annals of the rheumatic diseases}, 69\penalty0 (2):\penalty0 374--379, 2010.

\bibitem[Nie and Wager(2021)]{nie2021quasiRlearner}
Xinkun Nie and Stefan Wager.
\newblock Quasi-oracle estimation of heterogeneous treatment effects.
\newblock \emph{Biometrika}, 108\penalty0 (2):\penalty0 299--319, 2021.

\bibitem[Papadopoulos(2008)]{papadopoulos2008inductive}
Harris Papadopoulos.
\newblock Inductive conformal prediction: Theory and application to neural networks.
\newblock In \emph{Tools in artificial intelligence}. Citeseer, 2008.

\bibitem[Rokach and Maimon(2005)]{rokach2005top}
Lior Rokach and Oded Maimon.
\newblock Top-down induction of decision trees classifiers-a survey.
\newblock \emph{IEEE Transactions on Systems, Man, and Cybernetics, Part C (Applications and Reviews)}, 35\penalty0 (4):\penalty0 476--487, 2005.

\bibitem[Romano et~al.(2019)Romano, Patterson, and Candes]{romano2019conformalized}
Yaniv Romano, Evan Patterson, and Emmanuel Candes.
\newblock Conformalized quantile regression.
\newblock \emph{Advances in neural information processing systems}, 32, 2019.

\bibitem[Shafer and Vovk(2008)]{shafer2008tutorial}
Glenn Shafer and Vladimir Vovk.
\newblock A tutorial on conformal prediction.
\newblock \emph{Journal of Machine Learning Research}, 9\penalty0 (3), 2008.

\bibitem[Snapinn and Jiang(2007)]{snapinn2007responder}
Steven~M Snapinn and Qi~Jiang.
\newblock Responder analyses and the assessment of a clinically relevant treatment effect.
\newblock \emph{Trials}, 8:\penalty0 1--6, 2007.

\bibitem[Straube et~al.(2010)Straube, Derry, Moore, Paine, and McQuay]{straube2010pregabalinResponder}
Sebastian Straube, Sheena Derry, R~Andrew Moore, Jocelyn Paine, and Henry~J McQuay.
\newblock Pregabalin in fibromyalgia-responder analysis from individual patient data.
\newblock \emph{BMC Musculoskeletal Disorders}, 11:\penalty0 1--8, 2010.

\bibitem[Tibshirani et~al.(2019)Tibshirani, Foygel~Barber, Candes, and Ramdas]{tibshirani2019conformalweighted}
Ryan~J Tibshirani, Rina Foygel~Barber, Emmanuel Candes, and Aaditya Ramdas.
\newblock Conformal prediction under covariate shift.
\newblock \emph{Advances in neural information processing systems}, 32, 2019.

\bibitem[Vicente et~al.(2020)Vicente, Ballensiefen, and J{\"o}nsson]{vicente2020personalised}
Astrid~M Vicente, Wolfgang Ballensiefen, and Jan-Ingvar J{\"o}nsson.
\newblock How personalised medicine will transform healthcare by 2030: the icpermed vision.
\newblock \emph{Journal of Translational Medicine}, 18:\penalty0 1--4, 2020.

\bibitem[Wager and Athey(2018)]{wager2018estimationCF}
Stefan Wager and Susan Athey.
\newblock Estimation and inference of heterogeneous treatment effects using random forests.
\newblock \emph{Journal of the American Statistical Association}, 113\penalty0 (523):\penalty0 1228--1242, 2018.

\end{thebibliography}

\clearpage
\appendix

\section{Marginal vs Conditional Responder Testing} \label{appendix:example}

The test proposed for $H_{0, i}: (X_i, Y_i) \sim (X_0, Y_0)$, is a test of the joint distributions, as opposed to a test of the conditional distributions $H_{0,i}: Y_i(1) | X_i = x_i \sim Y_i(0)|X_i = x_i$. This implies, that the type I error is not controlled conditional on $X$, but rather marginalized over $X$, i.e., 
\begin{equation} \label{eq:joint_hypothesis}
\mathbb{P}_{(X_0, Y_0)| T = 1} (\mathrm{PV}_i \leq \alpha)  = \int \mathbb{P}_{X|T=1}(X = x) \mathbb{P}_{Y(0)| X=  x}\left(  \mathrm{PV}_{i} \leq \alpha) \right) dx \leq \alpha. \end{equation}  In most practical examples, the CARD's Type I error rate when testing conditional hypotheses is negligibly above the nominal level. However, it is possible to construct scenarios where the CARDs test's Type I error rate reaches up to 1 when testing these hypotheses.

Consider the case where $Y(0)|X = 0 + \epsilon$, $Y(1)| X = \ind(X > 0.01) + \epsilon$, and $\epsilon \sim N(0, 10^{-7})$. Furthermore, let $T \sim Ber(0.5)$ and $X \sim U[0, 1]$. 
For a very large number of observations, the discriminators will make the correct split on $X_1 > 0.01$, with a following split on $Y > 10^{-5}$ (or any other value which separates the distributions well). 
Due to these splits all of the null conditional hypotheses would be rejected (all observations for which $X_1 < 0.01$), yet, the joint type I error will be controlled at the expected level.

\section{Conformal Quantile Regression (CQR)}  \label{section:cqr}

To produce CQR, we split the data into a training set, $\mathcal{I}_{tr}$, and a calibration set, $\mathcal{I}_{cal}$. First, the training set is used for fitting a quantile regression model (such as linear quantile regression, quantile boosting, quantile trees, etc.). The model is trained to predict the conditional quantiles $q_{\alpha_l}(x)$ and $q_{\alpha_h}(x)$, where $\alpha_l + \alpha_h = \alpha$, and $1 - \alpha$ is the coverage probability.
Then, the calibration set is used to estimate the non-conformity score, $$ s(X_i, Y_i) = \max \left(q_{\alpha_l}(X_i) - Y_i, Y_i - q_{\alpha_h}(X_i) \right).$$ To construct the CI, the $ 1 - \alpha$ quantile of the non-conformity scores are computed on the calibration set, denoted by $Q_{1 - \alpha}$. Finally, the confidence interval is, $$ C(X_{n + 1}) = [q_{\alpha_l}(X_i) - Q_{1 - \alpha},  q_{\alpha_h}(X_i) + Q_{1 - \alpha}]. $$ To obtain p-values, we use the following, \begin{equation} 
    \mathrm{pv}_i = \frac{1 + \sum_{x_j, y_j \in \mathcal{I}_{cal}} \ind \left( s(x_j, y_j) \geq s(x_i, y_i) \right)}{|\mathcal{I}_{cal}| + 1}.
\end{equation}

\section{Global Testing} \label{appendix:global}

The task of detecting responders is essentially a task of detecting out-of-distribution observation based on the response. A simple approach would be to calculate the empirical p-values of $Y_i, T_i = 1$ observations based on the null sample, the p-values is, $$ 
    \mathrm{pv}^{\mathrm{global}}_i = 2 \min \left( \frac{1 + \sum_{j=1}^{n_0}  \ind \left( y_j 
    \geq y_i \right)}{n_0 + 1},  \frac{1 + \sum_{j=1}^{n_0}  \ind \left( y_j 
    \leq y_i \right)}{n_0 + 1} \right).
$$ A major drawback of the method is its inability to handle subjects which are responders conditional on some $X$. To illustrate the issue, we will use a somewhat contrived example, which exaggerates what otherwise is a very common phenomena. 
Suppose that $X \sim U[0,1]$, $$Y(0) = -5 \cdot \ind(x < 0.5) + 5 \cdot \ind(x > 0.5) + \epsilon,$$ 
and 
$$Y(1) = -5 \cdot \ind(x < 0.25) + 5 \cdot \ind (x >0.5) + \epsilon,$$ where $\epsilon \sim N(0, 1)$. 

\begin{figure}[ht]
    \centering
    \includegraphics[width=1\linewidth]{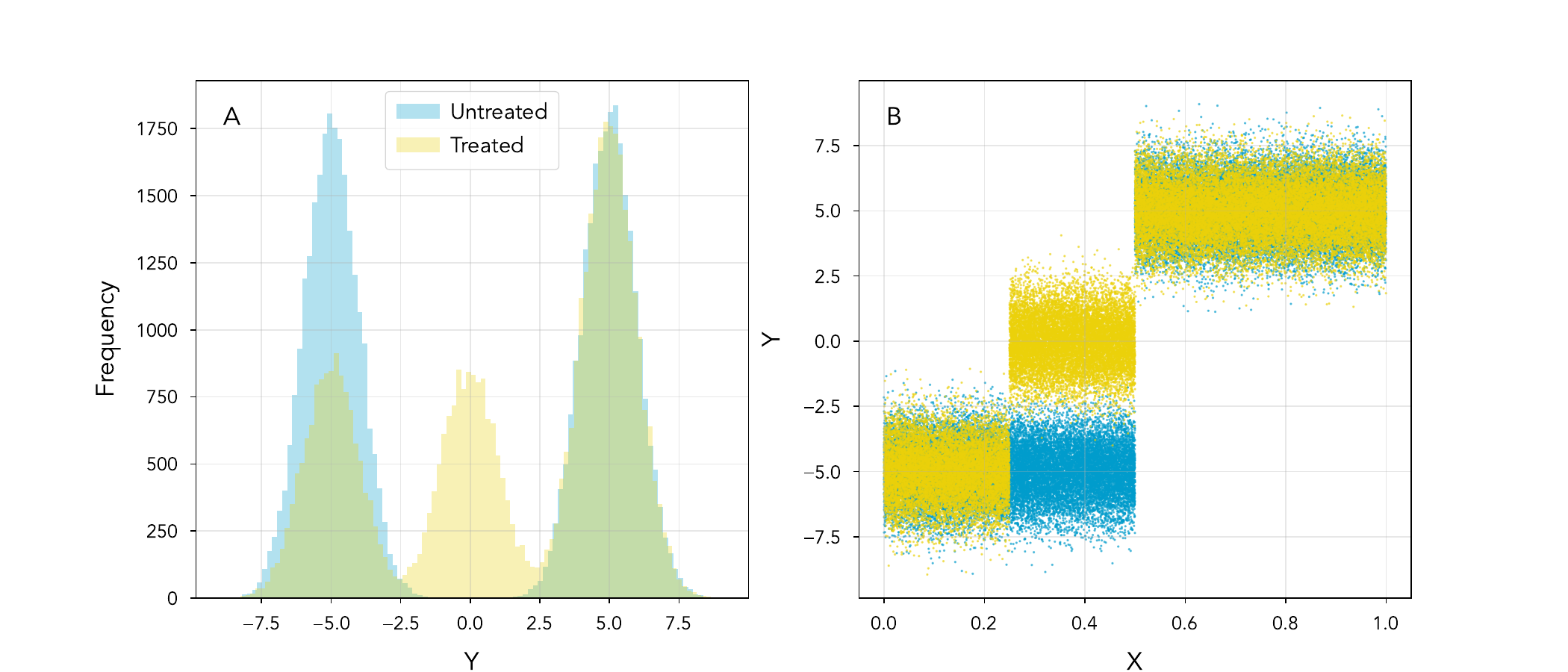}
    \caption{A. Histogram of $Y$, the responders are located between the two modalities of $Y(0)$. B. Scatter plot of $Y ~ X$, it is clear to see the responders around $Y = 0$ and $0.25< X<0.5$.}
    \label{fig:global_combined}
\end{figure}

The global testing approach has no power, since the responder do not appear extreme considering $Y$ (see \cref{fig:global_combined} A), while when considering $0.25<X < 0.75$, it is clear that the observation centered around $Y = 0$ are anomalies (see \cref{fig:global_combined} B). Indeed, in this example the power of the CARD is $1$, while the global testing has a power of $0$. 

\section{Proof of \texorpdfstring{\Cref{lemma:prop_adj}}{Lemma 1}} \label{appendix:proof}

\begin{proof}
    For simplicity, we assume that $\mathcal{I}_{ko}=\{1,\dots,k\}$ and $i=k+1$.
    
    Denote $Z_i = (X_i,Y_i)$, $z_i = (x_i,y_i)$.
    Let $c= \frac{\bb{P}(T=0)}{\bb{P}(T=1)}$, and let 
    $$\tilde{w}_i(x,y)=\begin{cases} 
    1 & 1\le i \le k \\
    c \frac{e(x)}{1-e(x)} & i= k+1
    \end{cases}.$$
    Under the null hypothesis, it holds that for every $x,y$, $1\le i_0\le k$
    \begin{equation}\label{eq:exchangeability}
        \bb{P}((X,Y)=(x,y)|T=0)\tilde{w}_{k+1}(x,y) = \bb{P}((X,Y)=(x,y)|T=1)\tilde{w}_{i_0}(x,y).
    \end{equation} 
    Indeed, the left hand side of \cref{eq:exchangeability} is 
    \begin{align*}
    &\frac{\bb{P}((X,Y)=(x,y)|T=0)\bb{P}(T=0)P(T=1|X=x)}{\bb{P}(T=1)P(T=0|X=x)} = \\
    &\frac{\bb{P}(T=0 | (X,Y)=(x,y)) \bb{P}((X,Y)=(x,y)) \bb{P}(T=1|X=x)}{\bb{P}(T=1)P(T=0|X=x)} = \\
    &\frac{\bb{P}((X,Y)=(x,y))\bb{P}(T=1|X=x)}{{\bb{P}(T=1)}} =    \\
    &\bb{P}((X,Y)=(x,y)|T=1).
    \end{align*}

    The first equality is by Bayes rule, the second equality is due to the null hypotheses ($\mathbb{P}(Y = y, T = t | X = x) = \mathbb{P}(Y = y | X = x) \mathbb{P}(T = t | X = x)$) and the last equality is by definition. 
    
    \Cref{eq:exchangeability} implies that $(Z_1,...,Z_{k+1})$ are weighted exchangeable (\cite[Definition~1]{tibshirani2019conformalweighted}) -- for every permutation $\sigma$,
    \begin{align*}
        \frac{\bb{P}(Z_1 = v_1 ,\dots, Z_{k+1}=v_{k+1})}{\tilde{w}_1(v_1)\cdots \tilde{w}_{k+1}(v_{k+1})}= \frac{\bb{P}(Z_1 = v_{\sigma(1)} ,\dots, Z_{k+1}=v_{\sigma(k+1)})}{\tilde{w}_1(v_{\sigma(1)})\cdots \tilde{w}_{k+1}(v_{\sigma(k+1)})}.
    \end{align*}

    Now, we let $E_z$ be the event that $\{Z_1,...,Z_{k+1}\}= \{z_1,...,z_{k+1}\}$. By the above, there is a constant $C_z$ such that  
    $$
    \bb{P}(Z_1 = v_{\sigma(1)} ,\dots, Z_{k+1}=v_{\sigma(k+1)}) = C_z \tilde{w}_1(v_1)\cdots \tilde{w}_{k+1}(v_{k+1}) = C_z \tilde{w}_{k+1}(v_{\sigma(k+1)}).
    $$

    Assuming for simplicity that all the values are different, using the same steps as in the proof of \cite[Lemma~3]{tibshirani2019conformalweighted}, it holds that
    \begin{align*}
    \bb{P}(Z_{k+1}=z_j|E_z) =& \frac{\sum_{\sigma:\sigma(k+1)=j}\bb{P}(Z_1 = v_{\sigma(1)} ,\dots, Z_{k+1}=v_{\sigma(k+1)})}{\sum_{\sigma}\bb{P}(Z_1 = v_{\sigma(1)} ,\dots, Z_{k+1}=v_{\sigma(k+1)})} \\
    =&\frac{k!\tilde{w}_{k+1}(z_j)}{\sum_{l=1}^{k+1}k!\tilde{w}_{k+1}(z_l)} = w_j^*.
    \end{align*}
    
    Finally, notice that the score $s$ is constant given $E_z$. This implies together with the above
    $$ \bb{P}(\mathrm{pv}^w_i < \alpha|E_z) \leq \alpha .
    $$
    And after marginalizing we get the required result.
\end{proof}
\clearpage
\section{Additional RCT Simulations} \label{appendix:rct_sim}

\subsection{Power Results}

Additional power results for the RCT setting simulations (\cref{sect:rct_sim}). The plots are illustrate the power as a function of the number of observations.  
The conclusions from the \cref{sect:rct_sim} are similar across the various scenarios. The CARD method outperforms all methods except in the positive signal and homoscedastic noise scenario.

\subsubsection{\texorpdfstring{$p =10, \rho = 0.9$}{p=10, rho = 0.9}}

\begin{figure}[ht]
    \centering
    \includegraphics[width=0.95\linewidth ]{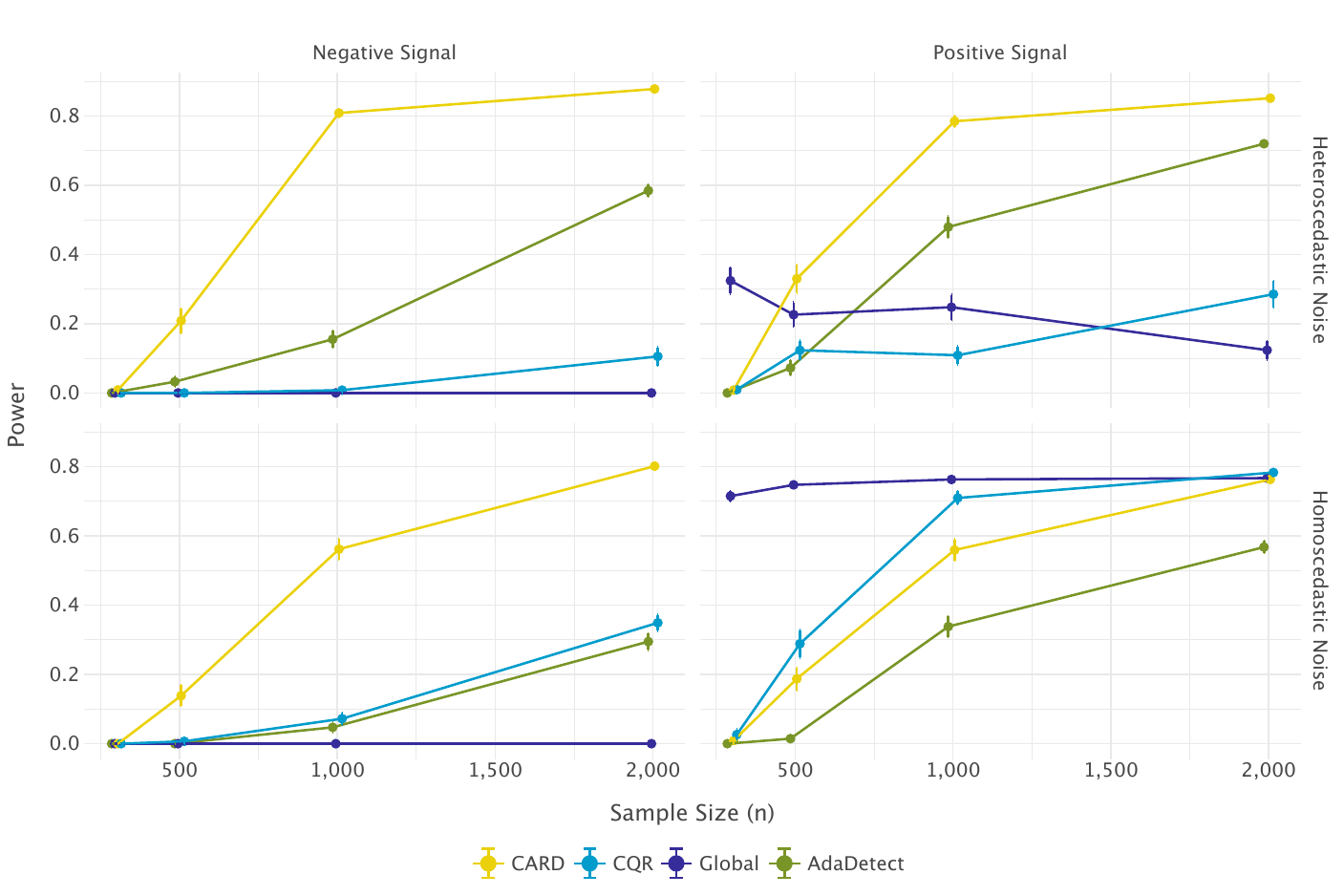}
    \caption{Power analysis of various methods for different sample sizes. The plots display the power as a function of sample size (n) in the ($p=10, \rho = 0.9$) scenario. Due to the correlations the AdaDetect method performs better compared to its performance when the covaraites are independent from one another, but still the best performing method is CARD except in the positive signal and homoscedastic noise, where the CQR and Global methods achieve better power for $n < 2000$.}
\end{figure}

\clearpage
\subsubsection{\texorpdfstring{$p =100, \rho = 0$}{p=100, rho = 0}}

\begin{figure}[ht]
    \centering
    \includegraphics[width=0.95\linewidth ]{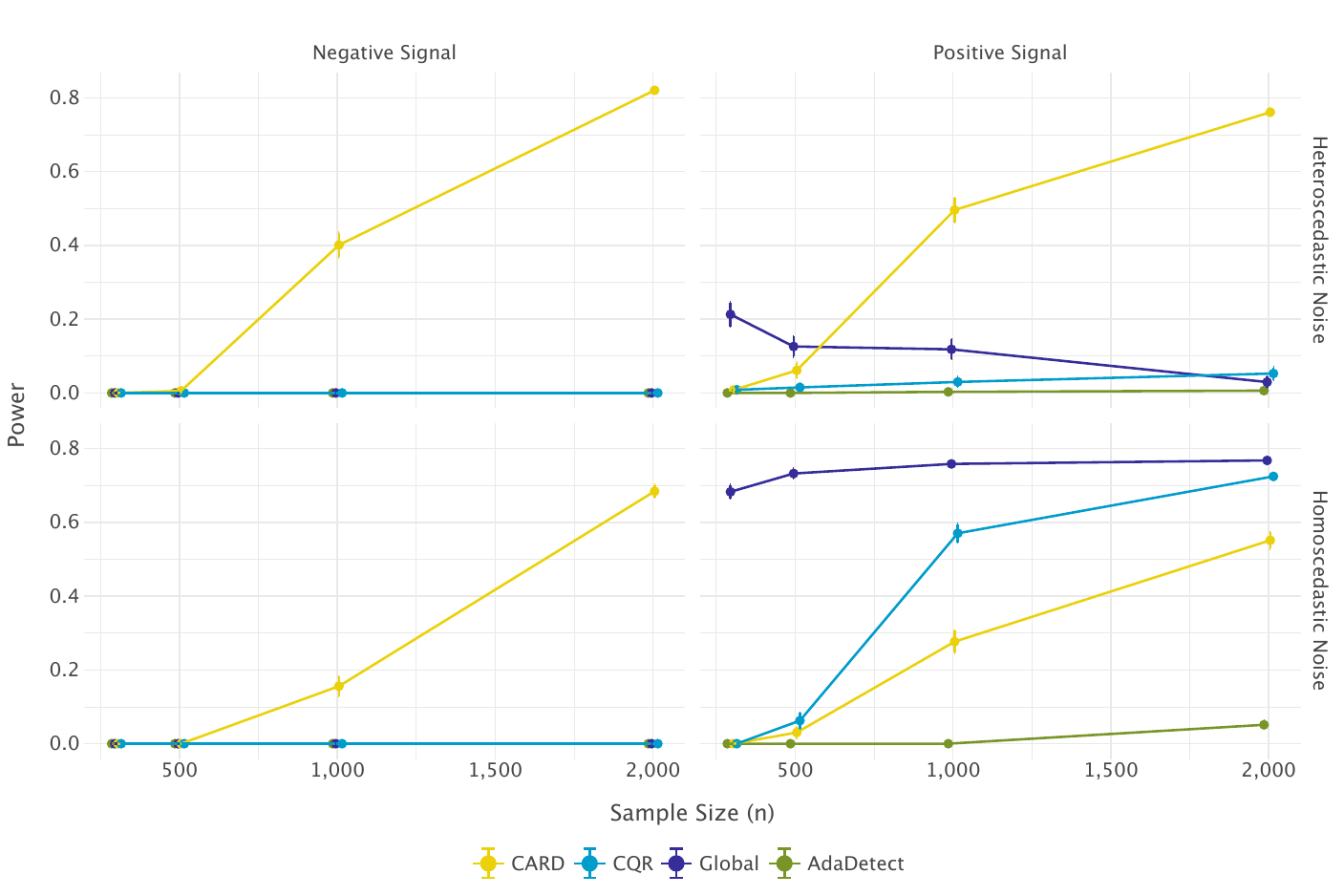}
    \caption{Power analysis of various methods for different sample sizes. The plots display the power as a function of sample size (n) in the ($p=100, \rho = 0$) scenario. The power of the CARD method is much higher in all scenarios except for the homoscedastic noise and positive signal. }
\end{figure}
\clearpage
\subsubsection{\texorpdfstring{$p =100, \rho = 0.9$}{p=100, rho = 0.9}}

\begin{figure}[ht]
    \centering
    \includegraphics[width=0.95\linewidth ]{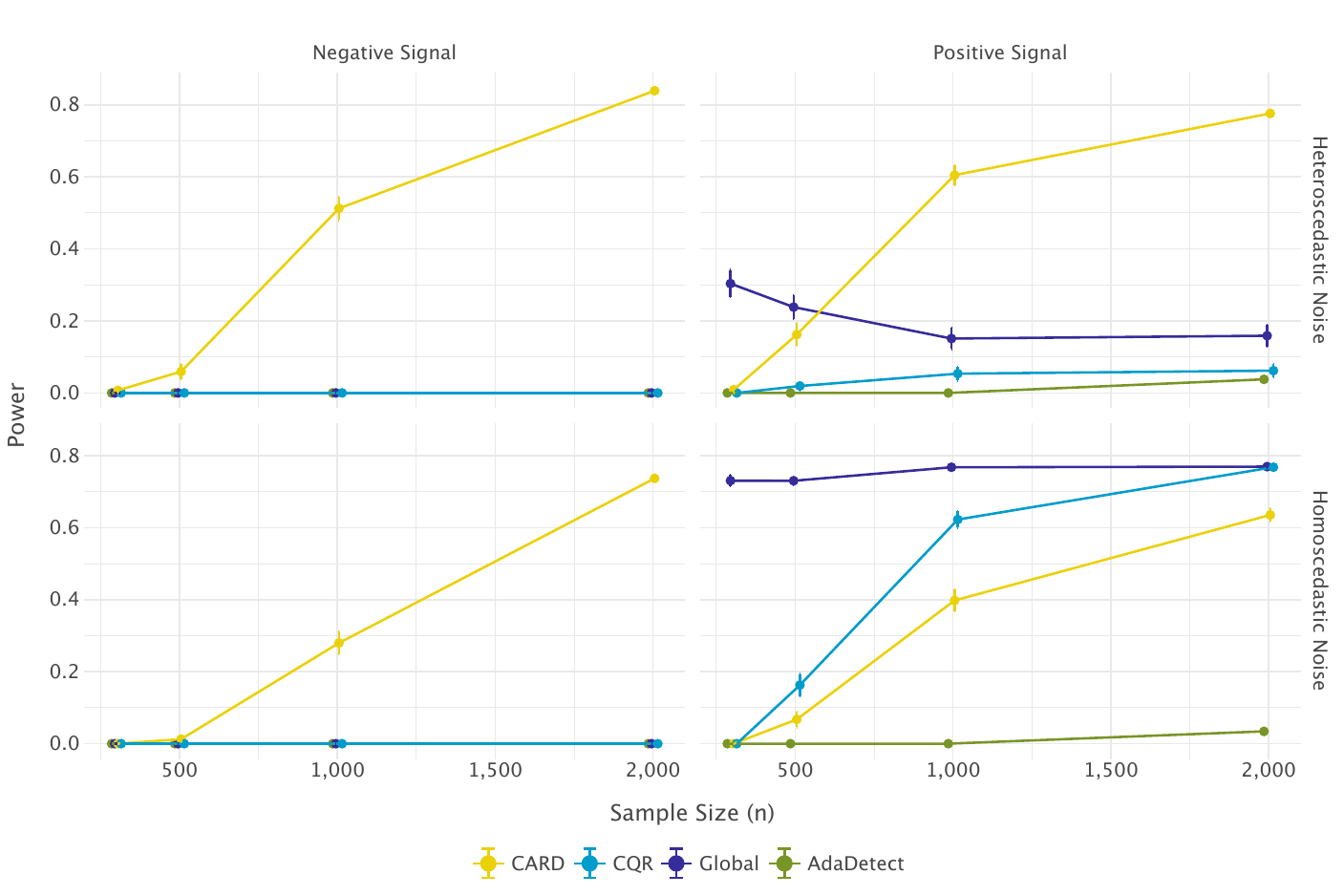}
    \caption{Power analysis of various methods for different sample sizes. The plots display the power as a function of sample size (n) in the ($p=100, \rho = 0.9$) scenario. The power of CARD method is much higher in all scenarios except for the homoscedastic noise and positive signal. }
\end{figure}

\clearpage
\subsection{FDR Results}

Across all simulations, all methods control the FDR at the expected level of $0.1$. 

\subsubsection{\texorpdfstring{$p =10, \rho = 0$}{p=10, rho = 0}}

\begin{figure}[ht]
    \centering
    \includegraphics[width=0.95\linewidth ]{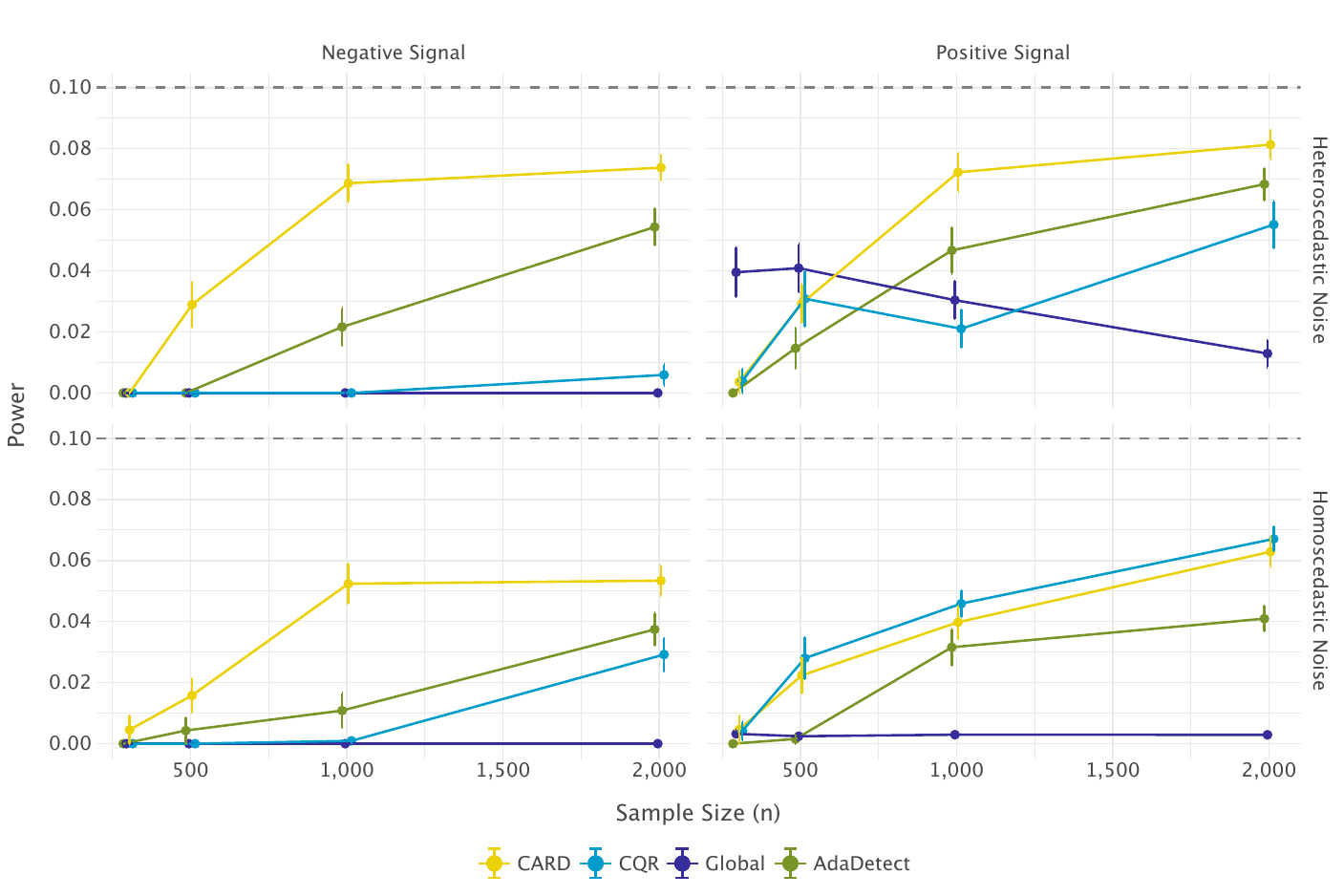}
    \caption{FDR of various methods for different sample sizes. The plots display the FDR as a function of sample size (n) in the ($p=10, \rho = 0$) scenario. }
\end{figure}

\clearpage
\subsubsection{\texorpdfstring{$p =10, \rho = 0.9$}{p=10, rho = 0.9}}

\begin{figure}[ht]
    \centering
    \includegraphics[width=0.95\linewidth ]{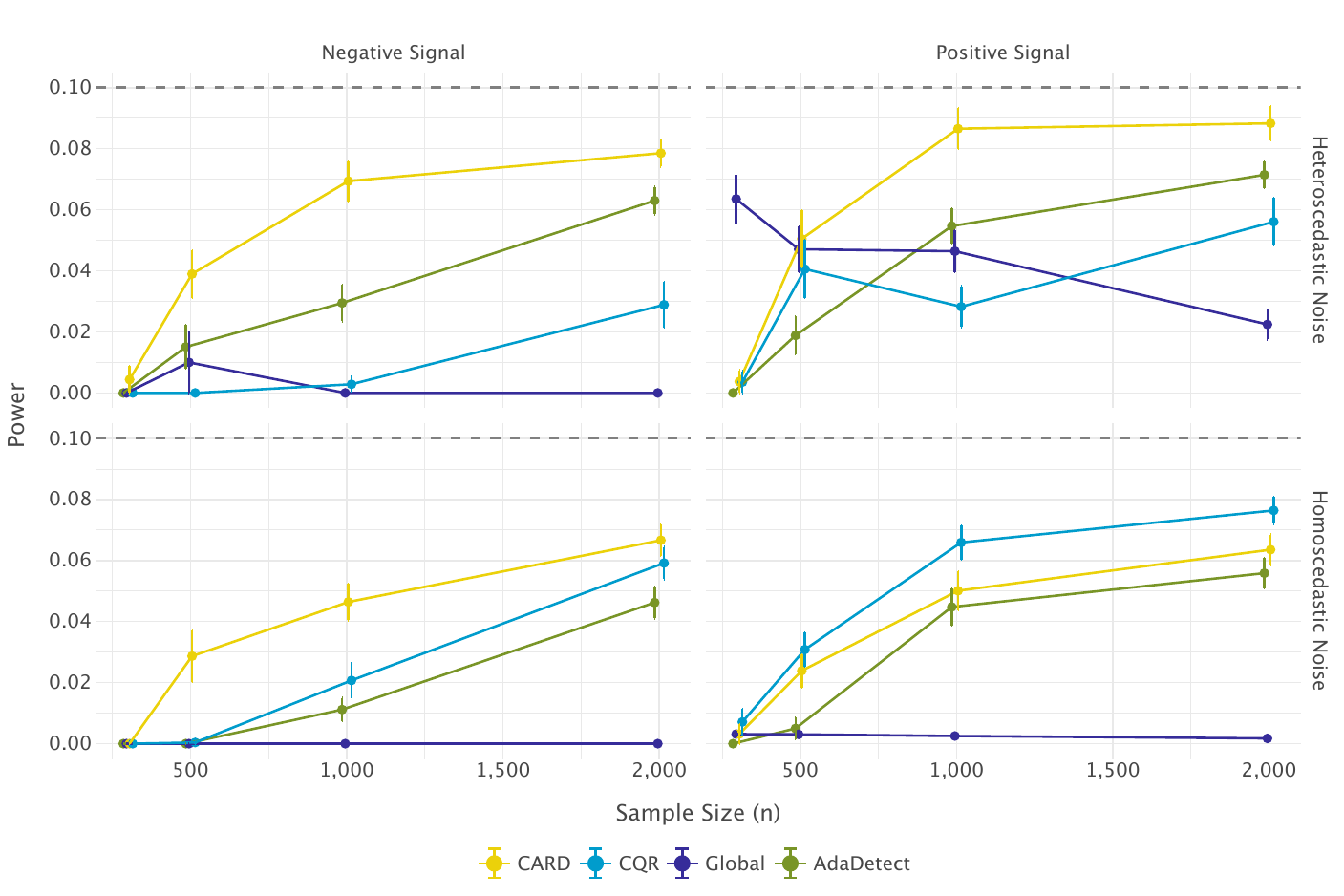}
    \caption{FDR of various methods for different sample sizes. The plots display the FDR as a function of sample size (n) in the ($p=10, \rho = 0.9$) scenario. }
\end{figure}
\clearpage
\subsubsection{\texorpdfstring{$p =100, \rho = 0$}{p=100, rho = 0}}

\begin{figure}[ht]
    \centering
    \includegraphics[width=0.95\linewidth ]{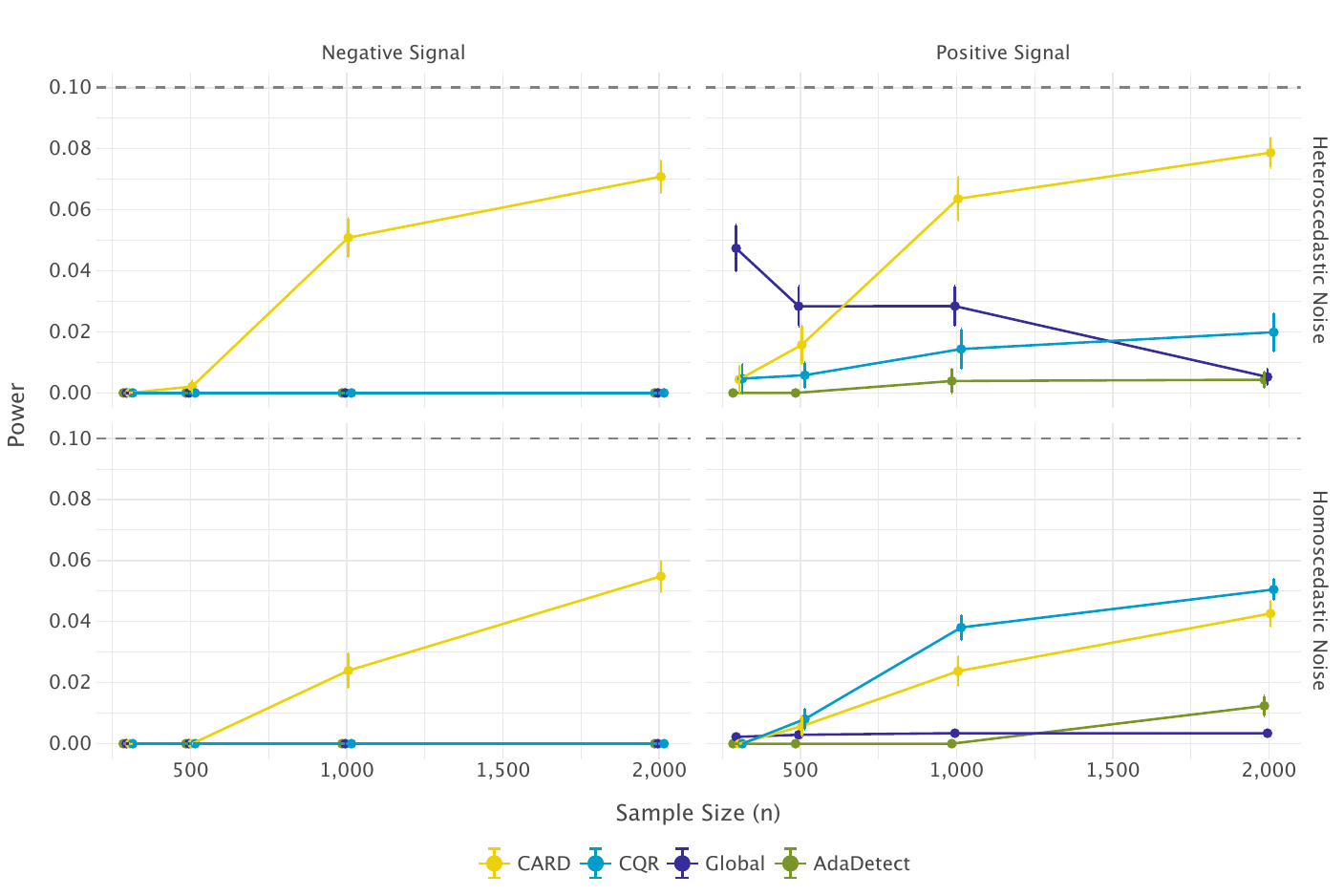}
    \caption{FDR of various methods for different sample sizes. The plots display the FDR as a function of sample size (n) in the ($p=100, \rho = 0$) scenario.
    }
\end{figure}
\clearpage
\subsubsection{\texorpdfstring{$p =100, \rho = 0.9$}{p=100, rho = 0.9}}

\begin{figure}[ht]
    \centering
    \includegraphics[width=0.95\linewidth ]{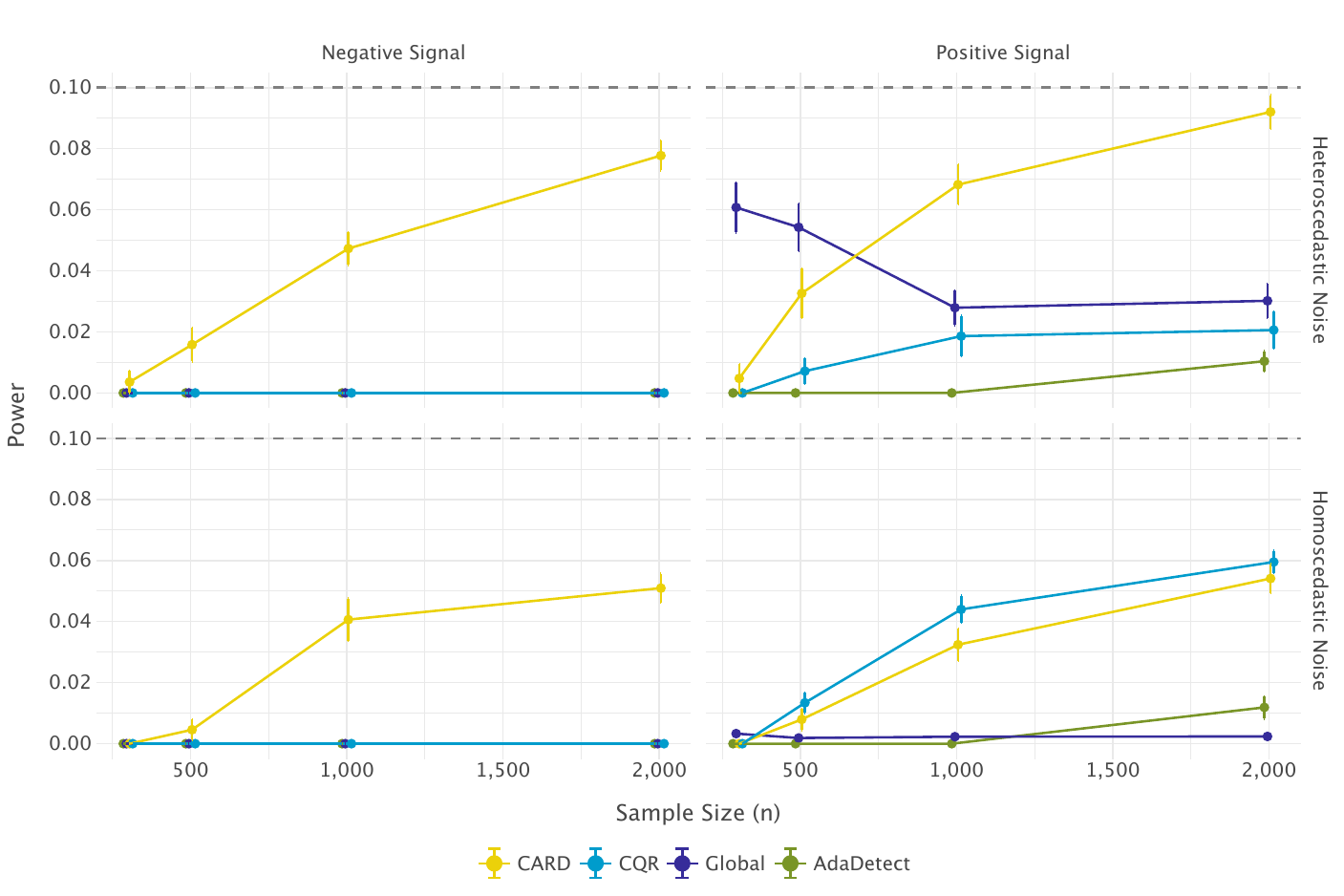}
    \caption{FDR of various methods for different sample sizes. The plots display the FDR as a function of sample size (n) in the ($p=100, \rho = 0.9$) scenario. }
\end{figure}

\end{document}